\documentclass[prc,aps,twocolumn,floatfix,showpacs,nofootinbib,superscriptaddress]{revtex4}
\usepackage{epsfig}
\newcommand{\etal}{\emph{et al.}}
\newcommand{\pnbar}{{|}}
\newcommand{\bd}{{\beta_2}}
\newcommand{\Sd}{{$^{32}$S}}
\newcommand{\Ars}{{$^{36}$Ar}}
\newcommand{\Are}{{$^{38}$Ar}}
\newcommand{\Caz}{{$^{40}$Ca}}
\newcommand{\Qzt}{{Q_{c2}^{(t)}(J,k)}}
\newcommand{\Qzs}{{Q_{c2}^{(s)}(J,k)}}

\newcommand{\bds}{{\beta_{c2}^{(s)}(J,k)}}
\newcommand{\bdt}{{\beta_{c2}^{(t)}(J,k)}}
\newcommand{\be}{\begin{equation}}
\newcommand{\ee}{\end{equation}}
\newcommand{\ba}{\begin{array}}
\newcommand{\ea}{\begin{array}}
\newcommand{\nn}{\nonumber}
\begin{document}

\title{Beyond--mean--field--model analysis of low-spin normal-deformed and 
       superdeformed collective states of \Sd, \Ars, \Are\ and \Caz}
\author{M. Bender}
\affiliation{Service de Physique Nucl\'eaire Th\'eorique,
             Universit\'e Libre de Bruxelles, C.P. 229, B-1050 Bruxelles,
             Belgium}
\author{H. Flocard}
\affiliation{CSNSM, Bt.104, F-91405 Orsay Campus, France}
\author{P.-H. Heenen}
\affiliation{Service de Physique Nucl\'eaire Th\'eorique,
             Universit\'e Libre de Bruxelles, C.P. 229, B-1050 Bruxelles,
             Belgium}
\date{May 8, 2003}
%
%
\begin{abstract}
We investigate the coexistence of spherical, deformed and superdeformed 
states at low spin in \Sd, \Ars, \Are\ and \Caz. The microscopic states are 
constructed by configuration mixing of BCS states projected on good 
particle number and angular momentum. The BCS states are themselves
obtained from Hartree-Fock BCS calculations using the Skyrme interaction 
SLy6 for the particle-hole channel, and a density-dependent contact force 
in the pairing channel. The same interaction is used within the 
Generator Coordinate Method to determine the configuration mixing 
and calculate the properties of even-spin states with positive parity. 
Our calculations underestimate moments of inertia. Nevertheless, 
for the four nuclei, the global structural properties of the states 
of normal deformation as well as the recently discovered superdeformed 
bands up to spin 6 are correctly reproduced with regard to both the 
energies and the transition rates.
\end{abstract}
\maketitle
%
%
\section{Introduction}
%
%
The existence of deformed bands in the spectrum of nuclei whose 
ground state is spherical has been established since the sixties 
for $^{16}$O, and since the early 1970s for $^{40}$Ca \cite{Nat76a}. 
With the modern large multi-detector $\gamma$ arrays like Euroball 
and Gammasphere, many more normal-deformed and even superdeformed 
rotational bands have been uncovered in these systems, such as those 
explored up to very high spin in doubly-magic $^{40}$Ca 
\cite{Ide01a,Chi03a}, and the adjacent transitional nuclei $^{36}$Ar 
\cite{Sve00a,Sve01a} and $^{38}$Ar \cite{Rud02a}. The occurence of
well-deformed prolate structures in such magic or close-to-magic nuclei
is understood as resulting from a drastic reorganization of the Fermi 
sea in which the oblate deformation driving last level of the shells
are emptied while orbitals originating from the $f_{7/2}$ shell are 
filled.
 
On the theoretical side, the new structures have been explored with the 
cranked mean-field (MF) method \cite{Ide01a} in which such a reorganization 
is naturally taken into account. On the other hand, thanks to conceptual 
and numerical progress that took place over the last decade, the shell 
model method is now also in position to analyze spectra in which spherical 
and well-deformed configurations coexist. However, in the latter case, 
the complexity of the calculation often prevents full scale diagonalizations 
in complete shells and specific choices must be made for the extension 
of the basis in order to keep computations within reach of present
technology. In some sense, prior to the diagonalization,
the expected physics is introduced in the Hamiltonian to
use. This choice is vindicated by the outcome the of calculations
and superdeformed bands built on particle excitations to
the $pf$ shells do come out. The cranked MF method does not have to
make such an \emph{a priori} choice; if a reorganization of
the Fermi sea is required, it will occur naturally as a
consequence of energy optimization. The present work illustrates 
a class of methods attempting to bridge the gap betwen these two 
approaches while remaining close to the MF in spirit. 
Indeed, the introduction of a diagonalization within a class of
MF collective states and the restoration of symmetries
(particle number, angular momentum) broken at the mean-field level
transfers the physical description from the intrinsic to the laboratory 
frame where the shell model naturally operates.
 
In the next section, we provide a quick overview of the qualitative 
features of our method, which is described in more detail in 
\cite{Val00a}. The necessary formalism and notations are also 
introduced. In the third section, we present our results for 
the four nuclei \Caz, \Are, \Ars\ and \Sd; a
selection which keeps in touch with recent experimental
progress. Moreover, taking into account an earlier
work on $^{16}$O \cite{Ben02a}, this choice allows us to test 
our method on a set of nuclei illustrating most
of the spectroscopic diversity of the $sd$ shell region of the
nuclear chart. In this section, our results are compared
with data and with those provided by shell model,
cranked MF, or other extended-MF calculations.
%
%
\section{The method}
\subsection{Effective Hamiltonian and Collective Hilbert Space}
The $N$-body physical states analyzed in this work are contained in 
the Hilbert space spanned by solutions $|\bd\rangle$ of Constrained 
Hartree-Fock-BCS (CHFBCS) equations \cite{BHR03}. In those equations, 
the constraint is imposed on the axial mass quadrupole moment 
operator ${\hat Q}_{20}$. The notation $\bd$ is a label
standing for any quantity in one to one correspondance 
with the expectation value \mbox{$Q_2=\langle\bd|{\hat Q}_{20}|\bd\rangle$}.
The single-particle wave functions from which BCS states are 
constructed are discretized on a three-dimensional mesh. As explained 
in Ref.\ \cite{Hee92a}, this technique provides accurate solutions of 
the mean-field equations. Note that our calculation does not assume
the existence of an inert core. 

As two-body interaction in the particle-hole channel of the Hamiltonian 
$\hat H$, we have chosen the SLy6 parametrization \cite{Cha98} of the 
Skyrme force. Most of our previous studies were performed with the SLy4
parametrization. Both sets have been fitted on the same set of 
observables, but differ by the way the center-of-mass 
motion is treated. In SLy4, only the diagonal part of the cm energy
is substracted from the total energy, while in SLy6, the full cm energy is
extracted self-consistently. This difference makes the calculations with
SLy6 more time consuming. The cm operator affects also the surface tension
of the Skyrme interaction and, in this respect, the SLy6 parametrization 
seems to be more satisfactory \cite{Ben00}. This better surface tension
is the motivation for the choice of SLy6 in the present study.
The pairing force is a zero-range, density-dependent force 
acting predominantly at the surface of the nucleus in order to 
describe the pairing effects in the particle-particle 
\mbox{$T=1$}, \mbox{$T_z=\pm1$} channel. The parameters of the 
latter force \cite{THF96} are identical for neutrons and protons 
and taken without readjustment from Refs.\ \cite{Val00a,Ben02a}. 
As in earlier calculations, for each nucleon species, the active 
pairing space is limited to an interval of 10 MeV centered at the 
Fermi surface. The present study does not therefore involve the 
definition of a new set of forces. It relies on well established 
interactions tested within the mean-field approach over a wide 
range of nuclei and phenomena covering the nuclear chart. 
This work is thus part of a program whose aim it is to perform 
an additional evaluation of this Hamiltonian by taking into
account the effects of quadrupole correlations.

For the sake of an easier comparison with the literature on quadrupole 
collective spectroscopy, we adopt the sharp edge liquid drop relation
to relate the  $\bd$ deformation parameter and the axial quadrupole 
moment $Q_2$
\be
\label{E01}
\bd = \sqrt{\frac{5}{16 \pi}} 
\, \frac{4 \pi Q_2}{3 R^2 A}\quad,
\ee
where the nuclear radius $R$ in fm at zero deformation is related 
to  its mass $A$ according to the standard formula 
\mbox{$R = 1.2 \, A^{1/3}$}. In this paper, only axial prolate and 
oblate deformations are considered. As we are mostly concerned with 
low-energy collective spectroscopy, we consider a range of 
values of $\bd$ covering all deformations such that the 
excitation energy of the constrained BCS states with respect to that of 
the spherical configuration \mbox{$|\bd=0\rangle$} is at most 20 MeV.

In the CHFBCS equations, we also introduce the correction terms 
associated with the Lipkin-Nogami prescription (see Ref.\ \cite{BHR03} 
for a description and for further references). Indeed, within schematic 
models, these terms have been shown to make the BCS solutions closer to 
those which would result from a full variation after projection on 
the particle number. Thus, they should be appropriate for 
calculations such as ours where a projection on $N$ and $Z$ 
is anyhow carried out at a later stage. Another benefit of the LN method
is that it suppresses the collapse of pairing correlations which may otherwise
occurs when the density of single-paricle states around the Fermi level
is small. It therefore ensures a smooth behavior as a function of the 
quadrupole moment of the overlap and energy kernels which intervene 
in the Generator Coordinate Method (GCM).

Several fundamental symmetries are broken in the BCS states $|\bd\rangle$,
which are eigenstates neither of the particle numbers nor of the  
angular momentum.\footnote{Among others that we do 
not discuss are translational, and isospin invariance.}
We restore these symmetries by means of a triple 
projection: first on the proton and neutron numbers $Z$ and $N$, then 
on the total spin $J$. For a given nucleus, the BCS state 
$|\bd\rangle$ has been determined with the usual constraints ensuring 
that the expectation values of the proton and neutron numbers 
have the correct $Z$ and $N$ values. In the following, we only 
select that component of the BCS state $|\bd\rangle$ which is 
an eigenstate of the particle number operators for 
the same values, discarding the $Z\pm 2,4,\ldots$ and $N\pm 2,4,\ldots$ 
components. For this reason, we do not introduce particle number 
labels and use the notation $\pnbar\bd\rangle$ for the particle-number 
projected state.

By contrast, the CHFBCS equations leading to the states $|\bd\rangle$
do not include any constraint on the angular momentum expectation value.
Starting from the $N$ and $Z$ projected states, we will consider 
separately all the collective Hilbert spaces spanned by the components 
$\pnbar J\,\bd\rangle$ resulting from a projection of the $\pnbar\bd\rangle$'s 
on the subspace of the even total angular momentum $J$. Hereafter,
the notation $\pnbar J,\bd\rangle$ stands for any of the states associated 
with the $2J+1$ values of the the third component of the angular 
momentum. Note that, because the initial CHFBCS equations
do not break the reflection or the time-reversal symmetry, the model 
provides no information on odd-spin states.

Finally, we diagonalize the Hamiltonian $\hat H$ within each of the 
collective subspaces of the non-orthogonal bases $\pnbar J,\bd \rangle$ 
by means of the Generator Coordinate Method (GCM) \cite{Hil53,Bon90,BHR03}.
This leads to a set of orthogonal collective states $\pnbar J,k\rangle$ 
where $k$ is a discrete index  which labels spin $J$ states 
according to increasing energy. As for $\pnbar J,\bd\rangle$ the 
notation $\pnbar J,k\rangle$ stands for any state of the spin multiplet.
A byproduct of the GCM are the collective wave functions
$g_{J,k}(\bd)$ describing the distribution of the states
$\pnbar J,k\rangle$ over the family $\pnbar J,\bd\rangle$. All 
collective properties discussed hereafter are directly computed 
from the $N$-body physical states $\pnbar J,k\rangle$.

We stress that the correlations 
introduced by the different configuration mixings of the initial 
mean-field wave functions $|\bd\rangle$ achieve several goals. First, 
the particle-number projection corrects a deficiency of the BCS description
of pairing in finite systems. Second, the angular momentum projection 
separates the dynamics according to spin and allows a direct calculation of
electromagnetic moments and transition probabilities in the laboratory 
rather than the intrinsic frame. Finally, performing a configuration
mixing over the coordinate $\bd$ by the GCM, we construct 
a set of orthonormal states $\pnbar J,k\rangle$ in which the large-amplitude 
quadrupole collective correlations are taken into account .
%
%
\subsection{Charge multipole and transition moments}
\label{subsect:moments}
The angular-momentum projection performs a transformation to 
the laboratory frame of reference and, therefore,
an intrinsic deformation cannot be unambiguously assigned to the 
projected states $\pnbar J,k\rangle$. For instance, all states 
$\pnbar 0,k\rangle$ have a quadrupole moment equal to zero. 
On the other hand, any multipole operator can be calculated in 
a straightforward, although sometimes tedious, manner directly 
from matrix elements involving the BCS states $|\bd\rangle$
 
It is, however useful to extract quantities analogous to intrinsic frame 
quantities from physically well-defined observables such as spectroscopic 
or transition moments, in order to achieve contact with standard 
mean-field approaches. For instance, when transition probabilities 
suggest that the two states $\pnbar J_i,k\rangle$ and $\pnbar J_f,k\rangle$ 
can be interpreted as forming a rotational band, one can define a 
charge intrinsic moment $\Qzt$ ($c$ stands for ``charge'' and $t$ for 
``transition'') from the $B(E2, J_i \to J_f)$ according to the 
standard formulas \cite{Lob75a}. Our model relies on the mixing
of purely axial quadrupole deformed configurations and describes 
\mbox{$K=0$} positive parity bands only.
Moreover, since the CHFBCS equations do not include a spin 
cranking term, the quality of the description of $\pnbar J,k\rangle$ 
states will deteriorate with spin $J$. In the following, we 
will consider only spins up to 10. A first possible definition 
of the intrinsic quadrupole moment $\Qzt$ is thus: 
\be
\label{E02}
\Qzt
= \sqrt{\frac{16 \pi}{5}  
  \frac{B(E2,J \to J-2)}
       {\langle J \,0 \, 2 \, 0 | J-2 \, 0 \rangle^2 e^2}}
,
\ee
whith the notation of Ref.\ \cite{CGC} for 
Clebsch-Gordan coefficients. Within the rigid rotor model, one
can also define an intrinsic quadrupole moment $\Qzs$ 
($s$ stands here for "spectroscopic") of a state with spin 
$J$ ($J\neq 0$) related with the spectroscopic quadrupole moment 
\be
\label{E03}
Q_c(J,k)
= \langle J,k\pnbar {\hat Q}_{c22}\pnbar J,k \rangle
,
\ee 
in the laboratory frame, via the relation
\be
\label{E04}
\Qzs 
= - \frac{2J+3}{J} \, Q_c(J,k)
.
\ee
In Eq.\ (\ref{E03}), it is understood that the bra and kets both 
correspond to the \mbox{$M=J$} component of the $2J+1$ multiplet. 
The values of $\Qzs$ and $\Qzt$ are equal when the rigid rotor 
assumption is strictly fulfilled. The differences between the values
of these two quantities will
tell us how well this assumption is satisfied by the calculated 
physical states. Instead of $\Qzs$ and $\Qzt$, we will equivalently 
consider the dimensionless quantities $\bds$ and $\bdt$ calculated 
according to a relation similar to Eq.\ (\ref{E01}) in which
the mass $A$ is replaced by the number of protons $Z$
since we are dealing with the charge quadrupole moment.

Technical details on the evaluation of $Q_c(J,k)$ and $B(E2)$ will
be given in an appendix below.
%
%
\section{Results}
%
%
\subsection{\Caz}
\begin{figure}[t!]
\epsfig{file=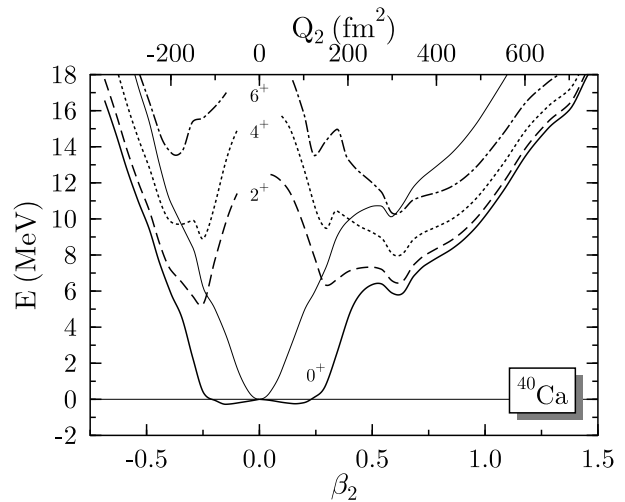}
\caption{\label{F01}
Nucleus \Caz; Projected energy curves versus the mass deformation 
($\bd$ and $Q_2$). The thin solid curve gives 
$\langle\bd\pnbar {\hat H}\pnbar\bd\rangle$ (MF), while the thick solid, 
dashed, dotted and dash-dotted curves correspond to
$\langle J,\bd\pnbar {\hat H}\pnbar J,\bd\rangle$ (PMF) for the 
values \mbox{$J=0$}, 2, 4 and 6 respectively. The energy origin is taken at
$\langle\bd=0\pnbar {\hat H}\pnbar\bd=0\rangle$.
}
\end{figure}
The ground state of the \mbox{$N=Z$} doubly-magic nucleus \Caz\ is 
spherical. As in several magic nuclei, the lowest excited
state is a $3^-$ state at 3.74 MeV. This state 
as well as other negative parity states is not included
in the Hilbert space of our calculation. Its study
requires parity breaking mean-field calculation involving octupole 
deformations as done in a previous study performed for Pb isotopes 
\cite{Hee01a}.

The lowest states of several even parity rotational bands are known 
since the early 70's \cite{Nat76a}. More recently, several normal- and 
also superdeformed bands have been identified up to very high spin 
thanks in particular, to a recent Gammasphere+Microball experiment 
\cite{Ide01a}.

\begin{figure}[t!]
\epsfig{file=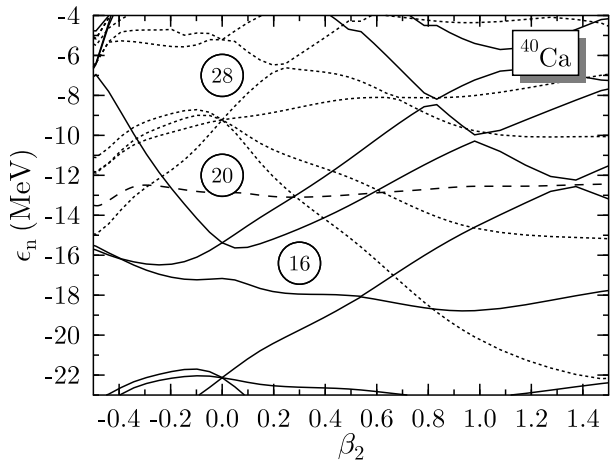}
\caption{\label{F02}
Nucleus \Caz; Self-Consistent HFBCS Nilsson diagram for neutrons. 
Except for an overall Coulomb shift, the proton  spectrum is almost 
identical. The dashed curve gives the Fermi energy.}
\end{figure}

In Fig.\ \ref{F01}, we have plotted the particle projected 
mean-field deformation energy curve (denoted below as MF) 
$\langle\bd\pnbar{\hat H}\pnbar\bd\rangle$  and the particle number 
and spin-projected mean-field curves 
$\langle J,\bd\pnbar{\hat H}\pnbar J,\bd\rangle$ (denoted below as PMF)
for low spin values. In addition to the expected well marked spherical
minimum, the MF curve displays a superdeformed secondary minimum
around \mbox{$\bd=0.5$}. The \mbox{$J=0$} PMF curve 
is almost flat over the range \mbox{$-0.2\leq\bd\leq 0.2$}. 
It also presents a secondary minimum around \mbox{$\bd=0.5$}. 
The minima of the higher spin PMF curves correspond to both oblate 
and prolate deformations at \mbox{$\bd \approx \pm0.2$} and to the
prolate superdeformation \mbox{$\bd=0.5$}. This latter minimum 
becomes the lowest one from spin 4 upward. These features of the 
mean-field curves are consistent with the HFBCS self-consistent 
single-particle diagram shown in Fig.\ \ref{F02}, where a small 
gap is visible at \mbox{$\bd=0.4$}. At \mbox{$\bd=0.6$},
a prolate deformation-driving orbital, originating from the $f_{7/2}$ 
shell, is occupied, creating a deformed shell gap.
Note that the deformed gap at \mbox{$\bd=0.9$} shows up as
a softening of the MF and PMF curves.
\begin{figure}[tbh!]
\epsfig{file=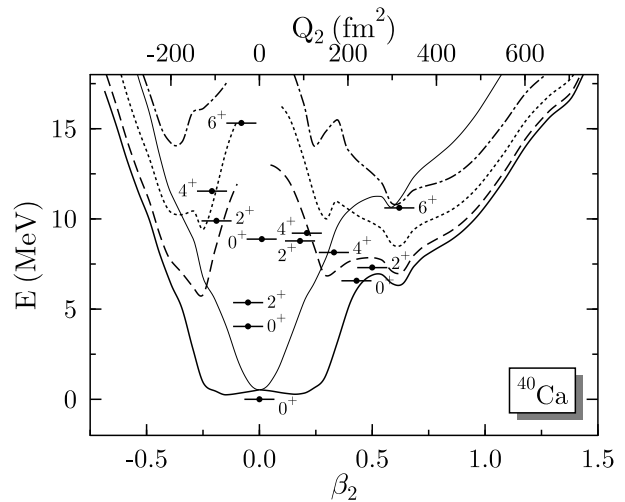}
\caption{\label{F03}
Nucleus \Caz; MF  $\langle\bd\pnbar{\hat H}\pnbar\bd\rangle$  (thin solid) 
and PMF $\langle J,\bd\pnbar{\hat H}\pnbar J,\bd\rangle$ (thick solid)
deformation energy curves. The ordinates of short horizontal segments 
give the energy $E_{J,k}$  of the GCM states (Eq.\ \ref{E05}). 
The abscissa of the black points indicate the 
mean deformation (${\bar\beta}_2$) of the
corresponding collective wave-function $g_{J,k}$ (Eq.\ \ref{E06}).  
The energy origin is taken at $E_{0,1}$.}
\end{figure}
\begin{figure}[tbh!]
\epsfig{file=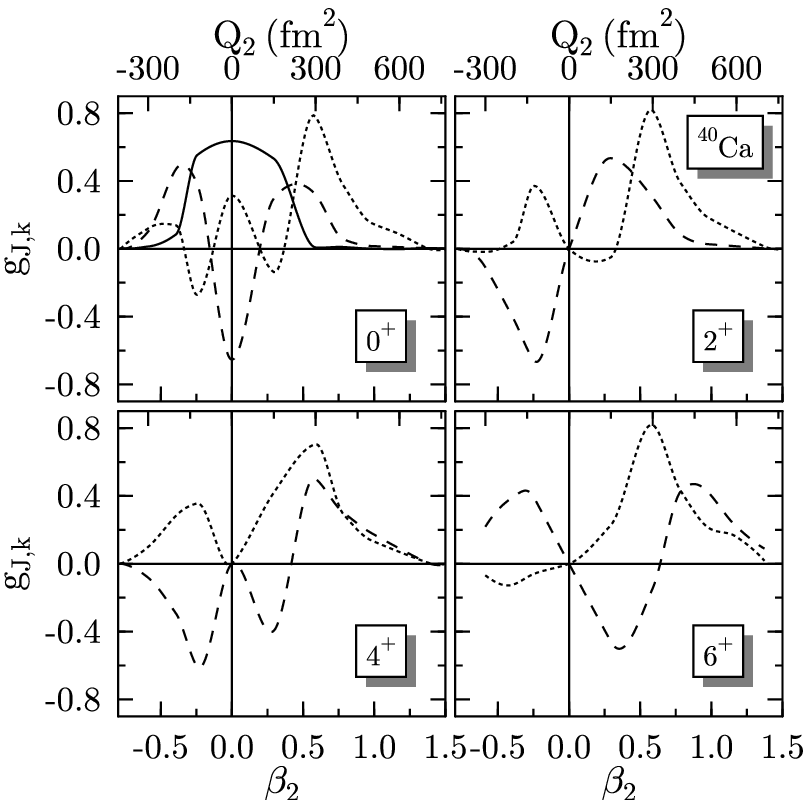}
\caption{\label{F05}
Collective GCM wave-functions $g_{J,k}$ for low-spin states of \Caz. 
The ground-state $0^+$ wave function is drawn with
a thick solid line. The wave functions of the ND and SD bands 
are drawn with dashed and dotted lines respectively.
}
\end{figure}
\begin{table*}[t!]
\caption{\label{T01}
Energy, spectroscopic moment and  transition amplitudes of
some of the ND (upper part)  and SD (lower part) bands
of \Caz. The $k$ labeling of the state 
refers to our calculation. The columns five to eight give 
the intrinsic moments, see Sect.\ \protect\ref{subsect:moments}. 
The last column gives the excitation energy of the
most likely corresponding experimental state \protect\cite{Ide01a}. 
}
\begin{tabular}{l|ccc|cccc|c}
\hline\noalign{\smallskip}
GCM               &
$E_{J,k}$         &
$Q_c$             &
$B(E2)\downarrow$ &
$Q_{c2}^{(s)}$    &
$\beta_{c2}^{(s)}$&
$Q_{c2}^{(t)}$    &
$\beta_{c2}^{(t)}$& 
$E_{exp}$        \\
State            &&& &&&& \\
$J_k^+$          &
(MeV)            &
($e$ fm$^2$)     &
($e^2$ fm$^4$)   &
($e$ fm$^2$)     &
                 &
($e$ fm$^2$)     &
                 &
(MeV)            \\
\noalign{\smallskip}\hline\noalign{\smallskip}
$0^+_2$ &  3.99 &   0.0 & --- &  --- &  ---   & ---  & ---  & 3.35 \\
$2^+_1$ &  5.40 &   2.2 & 112 & -7.8 & -0.031 & 75.2 & 0.30 & 3.91 \\
$4^+_2$ &  9.28 & -18.7 & 16  & 51.3 &  0.201 & 23.9 & 0.09 & 5.28 \\
$6^+_2$ & 13.39 & -34.3 & 187 & 85.8 &  0.336 & 77.4 & 0.26 & 6.93 \\
\noalign{\smallskip}\hline\noalign{\smallskip}
$0^+_3$  & 6.52 &   0.0 & --- & ---   &  ---- & ---  & ---  &  5.21 \\
$2^+_2$  & 7.30 & -38.2 & 447 & 133.9 & 0.525 & 150  & 0.60 &  5.63 \\
$4^+_1$  & 8.15 & -35.5 & 373 &  97.6 & 0.383 & 115  & 0.44 &  6.54 \\
$6^+_1$  &10.53 & -64.1 & 557 & 160.2 & 0.628 & 133  & 0.51 &  7.97 \\ 
$8^+_1$  &13.07 & -66.5 & 882 & 157.9 & 0.619 & 163  & 0.64 &  9.856 \\
$10^+_1$ &16.18 & -70.1 & 962 & 161.2 & 0.63  & 169  & 0.66 & 12.338 \\
\noalign{\smallskip}\hline
\end{tabular}
\end{table*}

In Fig.\ \ref{F03}, along with the MF and PMF curves, we show the 
excitation energies
\be
\label{E05} 
E_{J,k}=\langle J,k\pnbar{\hat H}\pnbar J,k\rangle
\ee
of the GCM states for the spins \mbox{$J=0$}, 2, 4 and 6. From
the GCM collective wave-function $g_{Jk}$ of each state 
$\pnbar J,k\rangle$, a mean deformation can be calculated
\be
\label{E06}
{\bar\bd}=\int \! \bd\;g_{J,k}(\bd)^2\,d\bd
.
\ee
This quantity does not necessarily coincide with the deformation 
$\bds$ calculated from Eqs.\ (\ref{E03}) of (\ref{E04}), but
provides a convenient indicator of the weight of deformed mean-field 
states in the projected state. It can be different from zero for 
\mbox{$J=0$} for which $\bds$ vanishes. In particular, it allows one
to detect the $0^+$ band head of a deformed band and to perform a band
classification of the states pending confirmation 
from the analysis of the in-band transitions. The magnitude of the 
axial quadrupole collective  correlations is given by the difference 
between the energies of the spherical configuration and of the GCM ground 
state, $E(0,1)$.
It is equal to 0.526 MeV, a value consistent with estimates in other
nuclei \cite{Hee01a}. Among the many $J_k^+$ GCM states, one 
observes at large deformation, \mbox{${\bar\bd}\approx 0.5$},
a subset ($0_3$, $2_2$, $4_1$, $6_1$) whose members can plausibly be
assigned to a superdeformed band (SD). This band is 
interpreted within the shell model as a 8p-8h band. The other excited 
states have smaller $\bar\bd$ values. At lower excitation energy, 
we will analyze whether the states ($0_2$, $2_1$, $4_2$, $6_2$) 
can be interpreted as the observed normal deformed (ND) band. 

In Fig.\ \ref{F05}, we have plotted the collective wave functions 
$g_{J,k}$. The position of their nodes can only be understood
if one keeps in mind the fact that, because the exchanges of 
intrinsic axes are included in a three-dimensional rotation, 
the quadrupole dynamics is not undimensional. It takes place
in the full quadrupole $\bd$, $\gamma$ plane along the six lines 
associated  with the gamma angles \mbox{$\gamma_j=2j\,\pi/6$}, 
\mbox{$j=0$}, \ldots, 5 \cite{BDF91}.  The underlying 
group structure also imposes that all non-zero-$J$ collective
wave-functions vanish at \mbox{$\bd=0$}. 
\begin{figure}[tbh!]
\epsfig{file=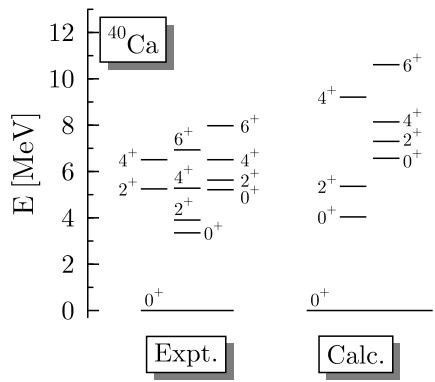}
\caption{\label{F04}
Comparison of experimental (left) and calculated (right) excitation 
spectrum of even spin and positive parity states in \Caz. 
}
\end{figure}

Fig.\ \ref{F04} compares the calculated and experimental excitation 
energies of the low spin states. The GCM states (right-hand side) can 
be assigned to ND or SD bands according to their $\bar\bd$ value. 
On the left side, we have drawn the
experimental states to which these bands most likely correspond.
In addition, at the extreme left side of the picture, we have 
plotted two other states observed within the same range of spins and 
excitation energies ($2^+$ at 5.249 MeV and $4^+$ at 6.509 MeV). 
However, we feel entitled to eliminate them from our discussion of the 
GCM results because, according to some theoretical models \cite{Ger69a},
they correspond to a \mbox{$K=2$} band which,  cannot be described 
within the present purely axial model.

The upper part of Table \ref{T01} collects data for the states 
of the ND band (labelled ``band 2'' in \cite{Ide01a}). They have already 
been  analyzed by Nathan \etal\ \cite{Nat76a} in the early 70s 
in terms of a 4p-4h structure. 

The GCM bandheads are at about the right excitation energy. 
However, the overall spectrum is too spread out. Since this 
is a general trend of this calculation, we defer a discussion of 
possible causes to the conclusions section. On the other hand, 
the $B(E2)$ values, or equivalently the $\Qzt$ moments, of the 
$E2$ transitions originating from the $2_1^+$ and the $6_2^+$
are in nice agreement with the experimental value 
\mbox{$\Qzt = 74 \pm 14$} $e$ fm$^2$ \cite{Ide01a}.
In our calculation, the transition down from the 
$4^+_2$ is reduced due to a coupling with the $4^+_1$ state of 
the SD band. This coupling is probably a spurious consequence 
of the too large spreading of the GCM ND band which pushes the
ND $4^+$ level in the vicinity of the SD $4^+$ state. Another 
consequence of the ND band spreading is that the SD $4^+$ and $6^+$ 
states are the lowest ones, while experiment suggests that this
band is not yrast below the highest observed spin (\mbox{$J=16$}). 
Our calculated value for the monopole transition matrix element 
\mbox{$|M(E0)| = 1.3$} $e$ fm$^2$ is smaller, but of the same order of 
magnitude as the experimental value $(2.6 \pm 0.1)$ $e$ fm$^2$ 
\cite{UBM77}. Note that no effective charge is used in our calculation. 

In the MF curve, the barrier between normal deformations and the SD 
minimum  does not exceed 0.5 MeV as seen in Fig.~\ref{F01}. The excitation
energy of this SD minimum is 10.1 MeV, a value much larger
than the value of 5.213 MeV observed for the SD bandhead
(band 1 in \cite{Ide01a}). Angular-momentum projection 
reduces this excitation energy to 6.1 MeV. Finally, the quadrupole
correlations due to configuration-mixing push the excitation energy 
of the SD bandhead to 6.52 MeV.  As seen in Fig.~\ref{F03}, this small
increase is related to the contribution of quadrupole correlations 
to the energy, which is larger for the $0_1^+$ ground state than for
the $0_3^+$ state. Thus, although the GCM result shows a significant 
improvement with respect to MF ones, the final outcome is not as 
satisfactory in \Caz\ as it is in $^{16}$O \cite{Ben02a}, where a  
better agreement with experiment had been found.

The GCM states ($0^+_3$, $2^+_2$, $4^+_1$, $6^+_1$) form a SD band 
with an average deformation \mbox{$\bar\bd=0.5$}. Their properties 
are given in the lower part of Table~\ref{T02}. As compared 
with experiment, see Fig.~\ref{F03}, the bandhead is slightly too 
high in excitation energy and 
the moment of inertia too small. Still, the agreement is much better
than for the ND band. One notes an irregularity for the $4^+$ state, 
due to the already mentioned mixing whith a nearby ND state. 
The calculated transition quadrupole moments for the various transitions 
in the SD band are in good agreement with the experimental value 
\mbox{$Q_{c2}^{(t)} = 180^{+39}_{-29}$} $e$ fm$^2$ given in 
Ref.\ \cite{Ide01a}, especially if one takes into account that in our 
calculation the $6_1^+\rightarrow 4_1^+$ and $4_1^+\rightarrow 2_2^+$ 
transition rates are decreased by a mixing of the $4_2^+$ level with 
the less deformed $4_1^+$ state. The proximity of the values of 
$Q_{c2}^{(t)}$ and $Q_{c2}^{(s)}$ shows also that the SD band satisfies 
the rotor criterion. 

A more recent analysis of transition quadrupole moments within the 
SD band presented in Ref.\ \cite{Chi03a} suggests that $Q_{c2}^{(t)}$
drops from values around 180 $e$ fm$^2$ at high spin to values around
120 $e$ fm$^2$ for the $6^+ \to 4^+$ transition. Our calculation is 
in agreement with this finding, although it predicts the change a few
units of angular momentum too low.
%
%
\subsubsection*{Particle-hole analysis of the GCM states}

\begin{table}[t!]
\caption{\label{T02}
Properties of the self-consistent $n$p-$n$h HF states and overlap
with low-lying $0^+$ GCM states. The second and tird columns give 
the excitation energy with respect to that of the (0p-0h) HF ground state
of the HF states ($E_{n\text{p-}n\text{h}}$) and of its \mbox{$J=0$}
component ($E_{J=0 n\text{p-}n\text{h}}$). The next two columns give the
the HF Intrinsic quadrupole moment $Q_2$ and the associated deformation 
$\bd$. The last three columns give the squared ovelaps
$|\langle J=0 k \pnbar n\text{p-}n\text{h} \rangle|^2$ with the lowest 
$\pnbar J=0, k \rangle$ GCM states. The normalization of the particle-hole 
states $|n\text{p-}n\text{h}\rangle$ is chosen such that the norm 
of its \mbox{$J=0$} component is equal to 1.
}
\begin{tabular}{l|cccc|ccc}
\hline\noalign{\smallskip}
state     & $E_{\text{HF}}$
          & $E_{J \text{HF}}$
          & $Q_2$
          & $\beta_2$
          & \multicolumn{3}{c}{$|\langle J k \pnbar J \text{HF} \rangle|^2$} \\
          & (MeV)
          & (MeV)
          & (fm$^2$)
          &
          & $0^+_1$ & $0^+_2$ & $0^+_3$ \\
\noalign{\smallskip}\hline\noalign{\smallskip}
     0p-0h &  0.002 &  0.000 &     0 &  0.000 & 0.6163 & 0.2170 & 0.0363 \\
  2p-2h(p) & 11.151 &  8.896 &    73 &  0.143 & 0.1189 & 0.0413 & 0.0330 \\
  2p-2h(n) & 11.462 &  9.071 &    73 &  0.143 & 0.1083 & 0.0421 & 0.0317 \\
     4p-4h & 11.767 &  8.729 &   211 &  0.414 & 0.0107 & 0.1131 & 0.0938 \\ 
  6p-6h(p) & 15.586 & 11.846 &   297 &  0.582 & 0.0000 & 0.0124 & 0.1626 \\
  6p-6h(n) & 15.791 & 12.002 &   298 &  0.584 & 0.0000 & 0.0110 & 0.1495 \\
     8p-8h & 13.247 &  8.732 &   396 &  0.777 & 0.0000 & 0.0043 & 0.1378 \\
   12p-12h & 31.625 & 26.565 &   667 &  1.308 & 0.0000 & 0.0000 & 0.0000 \\
\noalign{\smallskip}\hline
\end{tabular}
\end{table}

\begin{figure}[b!]
\epsfig{file=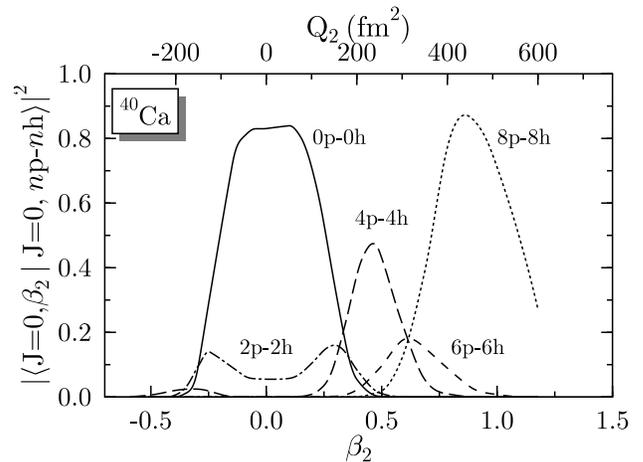}
\caption{\label{F06}
Square of the overlap $\langle J=0, \bd \pnbar n\text{p-}n\text{h} \rangle$ 
between the HF $|n\text{p-}n\text{h}\rangle$  states ($n=0$, 2, 4, 6, 8) and
the particle and spin \mbox{$J=0$} projected mean-field
states $\pnbar 0, \bd \rangle$. Because the two 2p-2h and the 6p-6h overlaps 
are very similar, only one for each is shown. The
normalization of a particle-hole state $|n\text{p-}n\text{h}\rangle$ 
is chosen so that the norm of its \mbox{$J=0$} component is equal to 1.
}
\end{figure}

The ND and SD bands are  often interpreted in terms of 4p-4h and 8p-8h 
configurations respectively. This is for instance the case in RMF 
calculations without pairing performed with the NL3 parametrization 
by the Lund group \cite{Sve00a}. Similarly, a shell model calculation by 
Caurier \etal\ \cite{Cau02a} in a restricted model space assigns the 
ND and SD bands to 4p-4h and 8p-8h configurations, respectively. It 
also finds that calculated transition probabilities related to quadrupole 
moments cluster around \mbox{$Q_{c2}^{(t)} = 170$} $e$ fm$^2$, and are 
slightly decreasing with angular momentum. It, thus, seemed of interest 
to perform a similar analysis of our results.

The definition of a particle-hole excitation is far from unambiguous.
For instance, in the shell model, it is defined with respect to the
spherical oscillator basis used in the calculation. However carefully 
this basis is chosen, it remains a technical rather than a physical 
reference. In keeping with the spirit of a mean-field method, we have 
defined the $n$p-$n$h states, $|n\text{p-}n\text{h}\rangle$, by requiring 
that, for each particle-hole configuration, they minimize the
Hartree-Fock energy with the same Skyrme Hamiltonian  used
in the projected GCM study. A deformed state $|n\text{p-}n\text{h}\rangle$, 
thus, differs from a particle-hole configuration built directly
by a simple redefinition of the orbital occupations in the spherical 
HF ground state $|0\text{p-}0\text{h}\rangle$. Since the definition 
of $|n\text{p-}n\text{h}\rangle$ does not involve the pairing channel,
all such states are Slater determinants and have the right particle 
number. In this way, we construct one state for 0p-0h, 4p-4h and 8p-8h 
configurations and two states for 2p-2h and 6p-6h ones. Because of 
the self-consistent redefinition of the orbitals, these states are 
not orthogonal to each other. Although their structure is very different,
it turns out that the energy of the spin \mbox{$J=0$} component extracted 
from the 2p-2h, 4p-4h, and 8p-8h states are almost degenerate, as can be seen 
in Table \ref{T02}. They are also very excited with respect to both the 
experimental and the GCM bandheads. There is no obvious explanation for 
this near-degeneracy, which was already noticed by Zheng \etal\ 
\cite{Zhe88a} in a much more schematic mean-field model.

The overlap between the particle-hole states and the \mbox{$J=0$} PMF 
states $\pnbar J=0,\bd\rangle$ is plotted in Fig.\ \ref{F06} as a function of
$\bd$. The overlap for the $|0\text{p-}0\text{h}\rangle$ is constant 
over the range $-0.2 \leq \bd\leq 0.2$. This range corresponds to a 
nearly constant PMF energy, as seen in Fig.\ \ref{F01}. The 2p-2h
overlap is small and vanishes outside this range. The 4p-4h overlap is 
gaussian shaped and peaked at \mbox{$\bd =0.5$}. The 8p-8h curve is peaked 
at an even larger deformation of $\bd=0.9$ and its maximum reaches a 
rather large value ($\approx 0.8$) indicating that the \mbox{$J=0$} 
component of the 8p-8h state is very close to to the 
$\pnbar J=0,\bd=0.9\rangle$ PMF solutions. All these different deformations 
are consistent with the locations of deformed shell gaps observed
in Fig.\ \ref{F02}. On the other hand, the overlap of the
8p-8h state is almost zero at the deformation $\bd\approx 0.5$
of the GCM SD band.

These results are already an indication that an analysis in terms  of pure 
particle-hole excitations can only provide a crude approximation. Another 
indication is that the energies of the $|n\text{p-}n\text{h}\rangle$
states are several MeV larger than those of the $\pnbar J=0,\bd\rangle$ states 
(Table \ref{T01}). This is confirmed by the overlaps given in the last 
three columns of Table \ref{T02}. It is seen that, with the exception
of the GCM ground state $0^+_1$, which, as expected, has a large overlap with
the spherical $|0\text{p-}0\text{h}\rangle$, neither the 4p-4h nor the 8p-8h 
HF solutions can be convincingly assigned to the ND or SD GCM bandheads. 
This is in line with findings of an earlier study of the light doubly-magic 
nucleus $^{16}$O \cite{Ben02a}.
%
%
\subsection{\Are} 
\begin{figure}[t!]
\epsfig{file=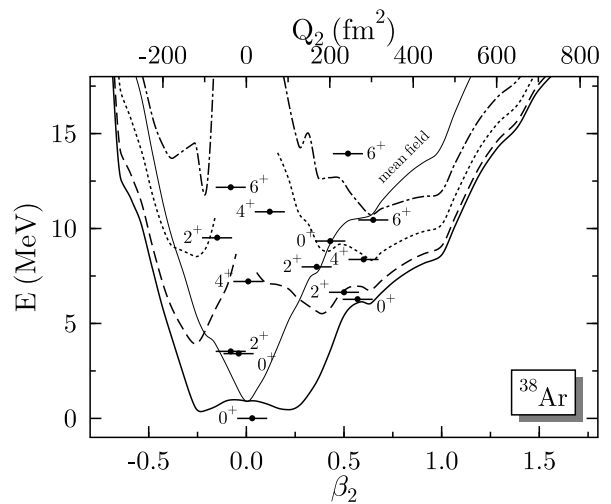}
\caption{\label{F07}
Nucleus \Are; Deformation energy curves $\langle\bd\pnbar{\hat H}\pnbar\bd\rangle$ 
(thin solid line) and $\langle J,\bd\pnbar{\hat H}\pnbar J,\bd\rangle$ with
\mbox{$J=0$}, 2, 4 and 6 corresponding to thick solid, dashed, dotted and 
dash-dotted lines, respectively. The ordinates of the short horizontal 
segments give the energy $E_{J,k}$ (Eq.\ \ref{E05}) of the GCM states. 
The abscissa of the black points indicates the mean deformation 
${\bar\beta}_2$ (Eq.\ \ref{E06}) of the GCM collective wave-function 
$g_{J,k}$. The energy origin is taken at $E_{0,1}$. 
}
\end{figure}

\begin{figure}[t!]
\epsfig{file=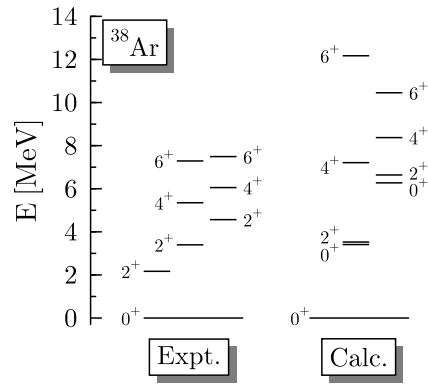}
\caption{\label{F08}
Same as Fig.\ \protect\ref{F04} for the nucleus \Are
}
\end{figure}
\begin{figure}[t!]
\epsfig{file=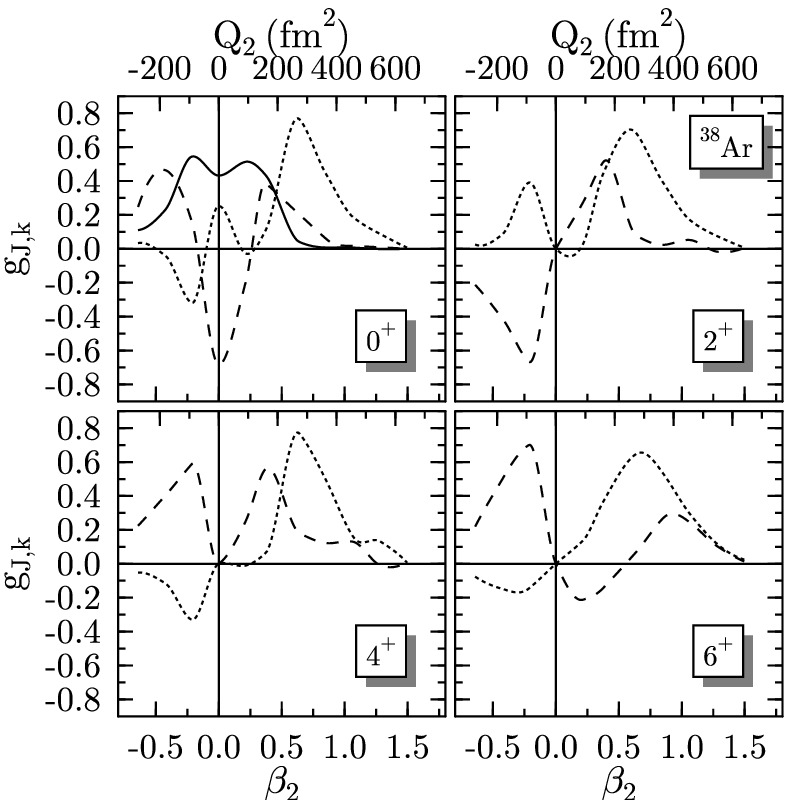}
\caption{\label{F09}
Collective GCM wave functions $g_{J,k}$ for the lowest spin states 
of \Are. The ground state wave function is drawn with
a thick solid line. The wave functions of the ND and SD bands are 
by dashed and dotted lines, respectively.}
\end{figure}

\begin{table*}[t!]
\caption{\label{T03}
Energy, spectroscopic moment and transitions amplitudes of
some of the ND band (upper part) and the SD band(lower part)
GCM states of \Are. The $k$ labeling of the state refers to 
our calculation. The columns five to eight give the intrinsic 
moments (see Sect.~\ref{subsect:moments}). The last column
gives the experimental excitation energy of the most likely 
state assigment \protect\cite{Rud02a}. 
}
\begin{tabular}{l|ccc|cccc|c}
\hline\noalign{\smallskip}
GCM               &
$E_{J,k}$         &
$Q_c$             &
$B(E2)\downarrow$ &
$Q_{c2}^{(s)}$    &
$\beta_{c2}^{(s)}$&
$Q_{c2}^{(t)}$    &
$\beta_{c2}^{(t)}$& 
$E_{exp}$        \\
state            & && &&&& \\
$J_k^+$          &
(MeV)            &
($e$ fm$^2$)     &
($e^2$ fm$^4$)   &
($e$ fm$^2$)     &
                 &
($e$ fm$^2$)     &
                 &
(MeV)            \\
\noalign{\smallskip}\hline\noalign{\smallskip}
$0^+_2$ &  3.44 &   0.0 & --- &  --- &  ---  & --- & ---  &      \\
$2^+_1$ &  3.61 &   6.9 & 104 & -24  & -0.11 & 72  & 0.33 & 3.94 \\
$4^+_1$ &  7.32 &   0.7 & 130 & -2   &  0.01 & 68  & 0.31 & 5.35 \\
$6^+_2$ & 12.41 &   8.3 &  47 & -21  &  0.09 & 39  & 0.18 & 7.29 \\
\noalign{\smallskip}\hline\noalign{\smallskip}
$0^+_3$ & 6.22 &   0.0 & --- & ---   & -- & --- & ---  &      \\
$2^+_2$ & 6.61 & -36   & 508 & 124 & 0.56 & 160 & 0.72 & 4.57 \\
$4^+_2$ & 8.11 & -52   & 611 & 142 & 0.64 & 147 & 0.62 & 6.05 \\
$6^+_1$ &10.34 & -59   & 611 & 146 & 0.66 & 140 & 0.63 & 7.49 \\ 
\noalign{\smallskip}\hline
\end{tabular}
\end{table*}
\begin{table}[t!]
\caption{\label{T04}
Out-of-band transitions and branching ratios for \Are.
The $k$ labeling of the state refers to our calculation.
Data are taken from \protect\cite{Rud02a}.
}
\begin{tabular}{c|ccc}
\hline\noalign{\smallskip}
$J_k^+ \to (J-2)_{k'}^+$  & $B(E2; J\to J-2)$
& \multicolumn{2}{c}{Branching Ratio}            \\
& ($e^2\,$fm$^4$)                            
&      Calc.        &   Exp.                      \\
\noalign{\smallskip}\hline\noalign{\smallskip}
$2^+_1 \to 0^+_1$ &  15 &       &                 \\
$4^+_1 \to 2^+_2$ &  22 &    14 &   5  (\emph{3}) \\
$6^+_2 \to 4^+_2$ & 126 &    73 &   5  (\emph{1}) \\
\noalign{\smallskip}\hline\noalign{\smallskip}
$2^+_2 \to 0^+_1$ & 1   &        &                \\
$2^+_2 \to 0^+_2$ & 1   &        &                \\
$4^+_2 \to 2^+_1$ & 3   &    0.5 &  40 (\emph{13})\\
$4^+_2 \to 2^+_3$ & 34  &        &                \\
$6^+_1 \to 4^+_1$ & 110 &    15  &  31 (\emph{6}) \\
\noalign{\smallskip}\hline
\end{tabular}
\end{table}

With the same techniques, we analyze three other nuclei, penetrating each
time somewhat deeper into the $sd$ shell. As for \Caz, a recent experiment 
\cite{Rud02a} has found high-spin states of \Are which can be interpreted 
in terms of bands, one of them corresponding to large deformation.
In both nuclei, the $0^+$ bandheads have not yet been seen.
A shell-model analysis has also been done in Ref.\ \cite{Rud02a}
within a restricted Hilbert space, as a full $s$-$d$, $f$-$p$ 
shell model calculation is presently beyond computational limits. 
Nevertheless, the overall quality of the agreement constitutes 
strong evidence of the importance of $f$-$p$ excitations. 
Since the \Caz\ subsection has shown that the real underlying structure is 
likely to be more complex than a single pure $m$p-$n$h excitation,
it is worth noting that our model has the ability to describe
coherent multiple excitation to higher shells.
 
Keeping in mind the limitations of the present calculation, we 
concentrate our analysis on the lower part of collective, even parity 
bands which, in Ref.\ \cite{Rud02a}, are labeled 1 and 2, respectively. 
For instance, our collective space  does not include two-quasiparticle
configurations and misses the lowest $2^+$ state which is well understood
in terms of a $\pi[d_{3/2}]^{-2}_{J=2}$ configuration \cite{WNe68,FHC75}. 
This assignement is compatible with  the pairing gap 
$\Delta_p$ at the proton Fermi surface that we obtain. 
Indeed, the excitation energy of this two-quasiparticle configuration is
approximated in perturbation by $2\Delta_p$, which is equal to 
1.96 MeV, rather close to the observed 2.17 MeV.

The MF and PMF energy curves of \Are\ presented in Fig.\ \ref{F07} 
are very reminiscent of \Caz. The MF curve exhibits a well-defined
spherical minimum. It is considerably softened by angular momentum 
projection, as shown by the  \mbox{$J=0$} PMF curve. Moreover, there 
is  a clear indication of a shell structure around \mbox{$\bd=0.5$}. 
At this deformation, the angular momentum projection on \mbox{$J=0$}
lowers the energy by approximately 5 MeV, reducing the excitation 
energy to slightly more than 6 MeV. The GCM results presented in
Figs.\ \ref{F08} and \ref{F09} are also similar to those obtained 
for \Caz: i) an isolated spherical ground state, ii) a set of states 
which can be grouped into a moderately deformed band (ND band) 
starting at 3.4 MeV and iii) a SD band with a 6.22 MeV bandhead. Keeping 
in mind that the present calculation predicts a too small moment of 
inertia, we estimate that the ND band starts at the correct energy 
while the SD band is located too high by about 1.8 MeV.
These results are comparable in quality to those reported of 
Ref.~\cite{Rud02a}, which however, in contrast to us, predict 
slightly too large moments of inertia.

The calculated in-band transition probabilities -- which are not yet 
measured --, the deformation and spectroscopic moments are given in 
Table \ref{T03}. They confirm the status of the SD band as a 
rotational band. 
The only existing  electromagnetic transition data
to be compared with our results concern branching ratios for several 
states of the two bands. Our results, are given in Table \ref{T04}.
%
%
\subsection{\Ars}
The low-energy spectrum of the \mbox{$N=Z$} nucleus \Ars\ is that of a 
transitional nucleus. In contrast to \Caz\ and \Are, a ND band is built
directly on the ground state \cite{End90aE}. Recently, a SD band has 
been observed up to a 16$^+$ state \cite{Sve00a,Sve01a}. 
The energy of the $0^+$ bandhead has been proposed to be at 4.3 MeV. 
Supporting data can be found in Ref.~\cite{Rop02a}. 
In the former references, the data are analyzed with the help of 
both shell model and cranked Nilsson-Strutinsky calculations. 
The shell model reproduces the ND band energies very well. 
Both methods explain the SD band in terms of the promotion of four 
particles to the $p$-$f$ shell. According to the diagram of Fig.~\ref{F02}, 
this excitation is associated with the deformed shell gap at 
\mbox{$\bd\approx 0.6$}. The calculations indicate that this 
SD band behaves as a good rotor. They also predict the position 
of the bandhead to better than 1 MeV. Both theoretical methods 
reach rather good agreement on $B(E2)$ values.  Results of similar quality
have been obtained by more schematic projected shell model calculations
however with appropriately adjusted parameters \cite{Lon01a}.

\begin{figure}[t!]
\epsfig{file=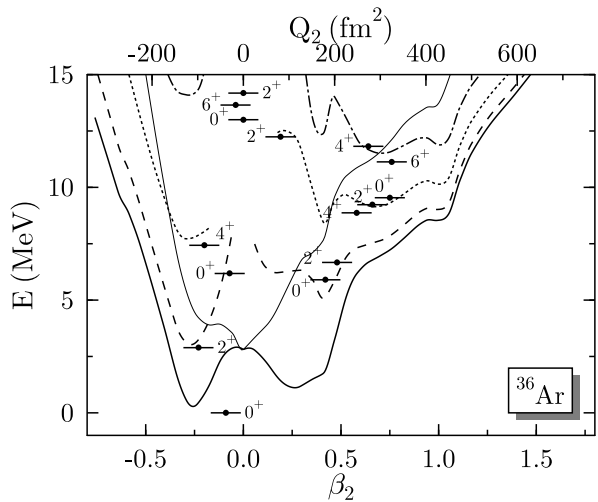}
\caption{\label{F10}
Same as Fig.\ref{F07}, but for the nucleus \Ars.
}
\end{figure}

\begin{figure}[t!]
\epsfig{file=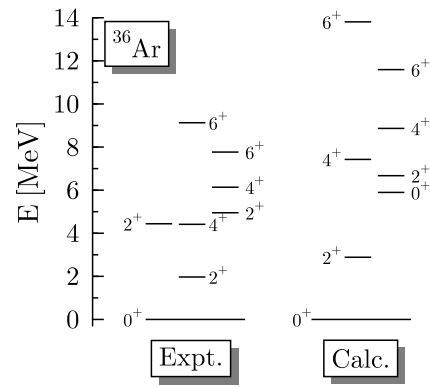}
\caption{\label{F11}
Same as Fig.~\protect\ref{F04} for the nucleus \Ars.
}
\end{figure}

\begin{figure}[b!]
\epsfig{file=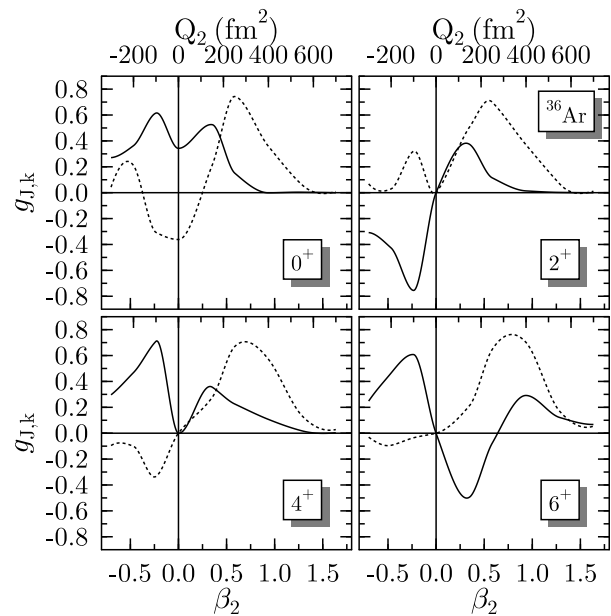}
\caption{\label{F12}
Collective GCM wave-functions $g_{J,k}$ for low-spin states of \Ars. 
The ground-state band and the SD band are drawn with solid and dotted 
lines respectively.
}
\end{figure}

As can be seen in Fig.\ \ref{F10}, the MF curve still presents a spherical 
minimum. It displays neither a secondary minimum nor even a plateau 
for a deformation close to \mbox{$\bd=0.6$}. The projection on spin 
markedly changes the picture. The ground state becomes oblate. At 
spin 0, the SD shell effect still does not show up as a minimum
while higher spin PMF curves present a well-defined
minimum which gradually shifts from \mbox{$\bd=0.45$} to $0.65$.
The properties of the GCM states, shown in Figs.\ \ref{F11} and \ref{F12},
qualitatively reproduce the observed features with one ND and one 
SD band. Note that the 4.4 MeV, $2^+$ level involves a two 
quasi-particle excitation and is, therefore, not included in our variational 
space. On the other hand, the SD bandhead energy is predicted too 
high by about 1.4 MeV and, as for the nuclei studied above, the moments
of inertia are too small. In Fig.\ \ref{F12}, one sees that the 
collective wave functions of the SD band are peaked around 
\mbox{$\bd\approx 0.55$}. This value is larger than that 
calculated in Ref.\ \cite{Sve00a,Sve01a}.

\begin{table*}[t!]
\caption{\label{T05}
Energy, spectroscopic moment and transition amplitudes of
ground band (upper part) and SD band (lower part)
GCM states of \Ars. The $k$ labeling of the state 
refers to our calculation. The columns five to eight give 
the intrinsic moments (see Sect.\ \protect\ref{subsect:moments}).
The last two columns give the experimental excitation energy and 
$B(E2)\downarrow_{\rm exp}$ corresponding to the most likely 
state assigment \protect\cite{Sve00a,Sve01a,Ram01a}. 
}
\begin{tabular}{l|ccc|cccc|cc}
\hline\noalign{\smallskip}
GCM               &
$E_{J,k}$         &
$Q_c$             &
$B(E2)\downarrow$ &
$Q_{c2}^{(s)}$    &
$\beta_{c2}^{(s)}$&
$Q_{c2}^{(t)}$    &
$\beta_{c2}^{(t)}$& 
$E_{exp}$         &
$B(E2)\downarrow_{exp}$\\
State            &&& &&&& \\
$J_k^+$          &
(MeV)            &
($e$ fm$^2$)     &
($e^2$ fm$^4$)   &
($e$ fm$^2$)     &
                 &
($e$ fm$^2$)     &
                 &
(MeV)            &
($e^2$ fm$^4$)   \\
\noalign{\smallskip}\hline\noalign{\smallskip}
$0^+_1$ &  0.00 &  0 & --- &  --- &  ---  & --- & --- & 0.00 &            \\
$2^+_1$ &  2.80 & 13 &  44 & -45  & -0.21 & 47  & 0.22& 1.97 & $60\pm 6$\\
$4^+_1$ &  7.43 & 12 & 103 & -34  & -0.16 & 60  & 0.28& 4.41 &         \\
$6^+_2$ & 13.65 & -1.3 &  93 & 3.3 & 0.02 & 55  & 0.26 & 9.18 \\
\noalign{\smallskip}\hline\noalign{\smallskip}
$0^+_2$ &  5.90 &   0 & --- &  --- &  ---  & --- & ---  & 4.33 &          \\
$2^+_2$ &  6.67 & -32 & 366 &  112 &  0.52 & 136 & 0.63 & 4.95 & \\
$4^+_2$ &  8.87 & -48 & 536 &  133 &  0.62 & 137 & 0.64 & 7.76 & $372\pm59$ \\
$6^+_1$ & 11.13 & -68 & 715 &  171 &  0.80 & 151 & 0.71 & 9.92 & $454 \pm 67$ \\
\noalign{\smallskip}\hline
\end{tabular}
\end{table*}

\begin{table}[t!]
\caption{\label{T06}
Out-of-band transitions in \Ars. The $k$ labeling of the states 
refers to our calculation. Data are taken from Refs.\ 
\protect\cite{Sve00a,Sve01a}.
}
\begin{tabular}{c|cc}
\hline\noalign{\smallskip}
$J_k^+ \to (J-2)_{k'}^+$ 
& \multicolumn{2}{c}{$B(E2)\downarrow (e^2\,$fm$^4)$}\\
&      Calc.        &   Exp.                      \\
\noalign{\smallskip}\hline\noalign{\smallskip}
$2^+_2 \to 0^+_1$ & 0.3   &     4.6$\pm$2.3        \\
$2^+_2 \to 0^+_3$ & 178   &                        \\
$4^+_2 \to 2^+_1$ & 3.3   &     2.5$\pm$0.4        \\
$6^+_1 \to 4^+_1$ & 57    &     5.3$\pm$0.8        \\  
\noalign{\smallskip}\hline
\end{tabular}
\end{table}

Table \ref{T05} confirms the oblate nature of the ND band, as can be
seen from the $g_{0,1}$ wave function and the rotor behavior of the 
SD band. This feature is not present in the data. From Tables \ref{T05} 
and \ref{T06}, one sees that the observed in-band $B(E2)$ values 
are slightly overestimated, while the out-of-band transition rates
are correctly reproduced. Our results show that the inclusion of 
collective quadrupole correlations supports the results obtained 
by more schematic models such as the cranked Nilsson calculations
presented in Refs.\ \cite{Sve00a,Sve01a} 
%
%
\subsection{$^{32}$S}
\begin{figure}[t!]
\epsfig{file=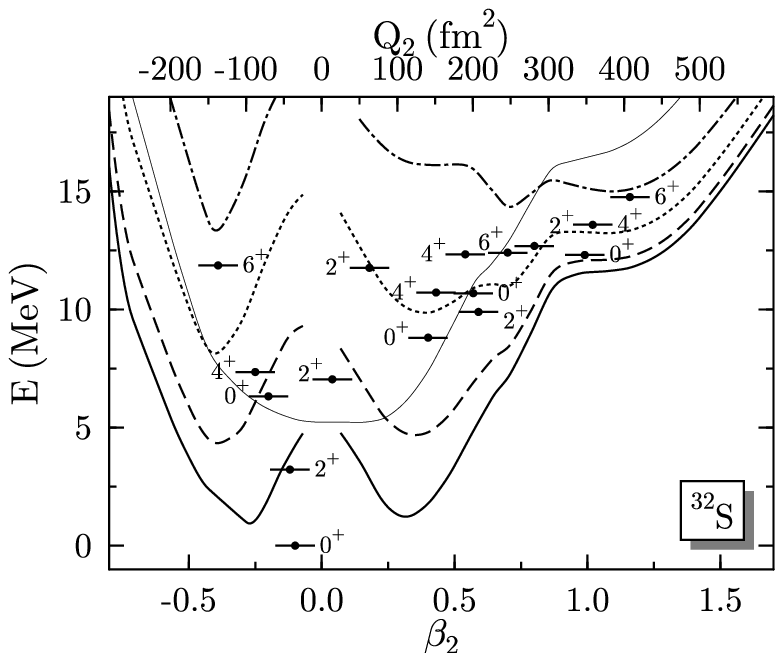}
\caption{\label{F13}
Same as Fig.~\ref{F07}, but for \Sd.
}
\end{figure}
\begin{figure}[t!]
\epsfig{file=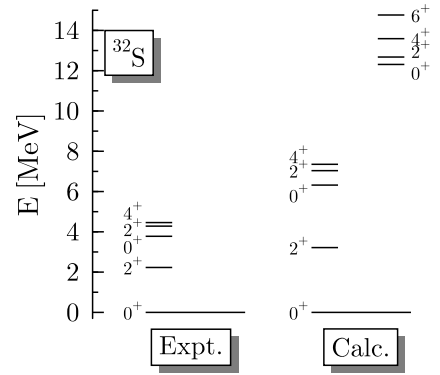}
\caption{\label{F14}
Same as Fig.~\ref{F04} for the nucleus \Sd.
}
\end{figure}
\begin{figure}[t!]
\epsfig{file=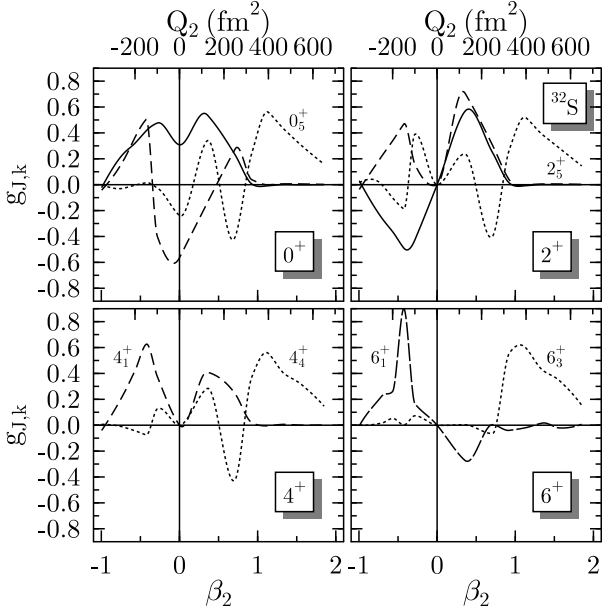}
\caption{\label{F15}
Collective GCM wave-functions  $g_{J,k}$ for the lowest spin 
states of \Sd. The first $0^+$ (ground state) and $2^+$ are drawn 
with a thick solid line. The dashed line corresponds to the $0^+$, $2^+$ 
and $4^+$ members of the two-phonon triplet, while the SD rotational 
states are drawn with a dotted lines respectively.
}
\end{figure}
\begin{table*}[t!]
\caption{\label{T07}
Data for the low-lying states in \Sd\ which can be interpreted 
as anharmonic vibrator states. The $k$ labeling of the states refers to 
our calculations.  Experimental data are taken from 
\protect\cite{Kan98aE} and \cite{Bab02aE}. 
}
\begin{tabular}{l|cccc|cc|cc}
\hline\noalign{\smallskip}    
$J_k^+$                  &
$E$ (MeV)                &
$Q_s$                    &
Transition               &
$B(E2)$                  &
$Q_{c2}^{(s)}$           &
$Q_{c2}^{(t)}$           &
$E_{exp}$                &
$B(E2)_{exp}$            \\
               &
(MeV)          &
($e$ fm$^2$)   &
               &
($e^2$ fm$^4$) &
($e$ fm$^2$)   &
($e$ fm$^2$)   &
(MeV)          &
($e^2$ fm$^4$) \\
\noalign{\smallskip}\hline\noalign{\smallskip}     
$0^+_1$ & 0.00 & 0.0  &                & --- & --- & ---  & 0.00 &          \\
$2^+_1$ & 3.22 &  2.3 &$2^+_1\to 0^+_1$&  38 & 2.3 & 43   & 2.23 & 61 $\pm$3\\
$0^+_2$ & 6.32 &  --- &$0^+_2\to 2^+_1$& 144 & --- & 12   & 3.78 & 72       \\ 
$2^+_2$ & 7.04 & -0.7 &$2^+_2\to 2^+_1$& 157 & -0.7 & 74   & 4.28 & 54 $\pm$3\\
        &      &      &$2^+_2\to 0^+_1$& 0.02 &    & 0.8  &      & 11 $\pm$2\\
        &      &      &$2^+_2\to 0^+_2$& 2.8 &     & 12   &      &          \\
$4^+_1$ & 7.35 & 11.7 &$4^+_1\to 2^+_1$& 94  & 12  & 58   & 4.46 & 72       \\
\noalign{\smallskip}\hline
\end{tabular}
\end{table*}
\begin{table}[t!]
\caption{\label{T08}
Energy, spectroscopic quadrupole moment and $E2$ transition amplitudes 
of the SD band states of \Sd. The $k$ labeling of the state refers to 
our calculation. The columns five to eight give the intrinsic 
moments (see Sect.\ \protect\ref{subsect:moments}).
}
\begin{tabular}{l|ccc|cccc}
\hline\noalign{\smallskip}
$J_k^+$           &
$E_{J,k}$         &
$Q_c$             &
$B(E2)\downarrow$ &
$Q_{c2}^{(s)}$    &
$\beta_{c2}^{(s)}$&
$Q_{c2}^{(t)}$    &
$\beta_{c2}^{(t)}$\\
                 &
(MeV)            &
($e$ fm$^2$)     &
($e^2$ fm$^4$)   &
($e$ fm$^2$)     &
                 &
($e$ fm$^2$)     &
                 \\
\noalign{\smallskip}\hline\noalign{\smallskip}
$0^+_5$ &  12.3 &  0  & ---  & --- & ---  & --- & ---  \\
$2^+_5$ &  12.7 & -39 &  719 & 136 & 0.77 & 190 & 1.08 \\
$4^+_4$ &  13.6 & -64 &  767 & 176 & 1.00 & 164 & 0.94 \\
$6^+_3$ &  14.8 & -82 & 1057 & 206 & 1.17 & 184 & 1.05 \\
\noalign{\smallskip}\hline
\end{tabular}
\end{table}

The MF energy curve for the mid-shell, \mbox{$N=Z$}, nucleus \Sd,
presented in Fig.~\ref{F13}, diplays a flat spherical minimum.
As can be expected from the single-particle spectrum diagram of 
Fig.~\ref{F02}, it does not show any structure around 
\mbox{$\bd \approx 0.5$}. Note that it presents an inflexion point at very 
large deformation, \mbox{$\bd\approx 1$}. In contrast, the \mbox{$J=0$} 
angular momentum projected curve exhibits two well-defined minima at small
deformation, the oblate one being the lowest. One also notes a 
stabilization of the SD minimum which, however,
never becomes yrast. These results are similar to those of 
other mean-field approaches \cite{Mol00a,Rod00a,Yam00a,Tan01a}.
Note that the authors of \cite{Tan01a} suggest that the SD 
structure may be stabilized at high spin by non-axial octupole
deformations.

The GCM calculation leads to the results displayed in Figs.~\ref{F14}
and \ref{F15}, and the moments and transition rates given in
Tables \ref{T07} and \ref{T08}. In agreement with experiment, the 
levels and energies are those of an anharmonic vibrator with 
a characteristic $0^+_1$, $2^+_1$ and a ($4^+_1$,$2^+_2$,$0^+_2$)
sequence. The vibrational pattern is slightly more prominent in 
the calculated ratios of $B(E2)$ values than in the observed ones. 
This may be due to an innacuracy in the description of the $2_1^+$ 
state which is almost spherical in our calculation while the
data suggest a strong prolate deformation. In comparison with 
similar calculations with the Gogny force \cite{Rod00a}, the SLy6 
force appears to be less performant on this specific point. On the 
other hand, the overall pattern and, in particular, the triplet of 
two-phonon states is better described by our calculation.     

The GCM predicts a SD band with a large deformation
\mbox{$\bd\approx 0.9$}, starting at the high energy of 12 MeV.
However, this energy must probably be scaled down, presumably 
by the same factor than the one that can be deduced from a comparison 
of experimental and GCM spectra on \Caz, \Are, \Ars\ and \Sd.
%
%
\section{Discussion and Conclusions}
This article belongs to a series exploring a model starting from 
a mean-field description of nuclear structure 
and an effective Hamiltonian valid for the entire nuclear chart.
The model attempts a global description of the nuclear ground states 
and low-energy spectra. Here, the SLy6 Skyrme interaction 
and a zero-range density dependent pairing force have been used. 
The building blocks are self-consistent $N$-body states, based on 
the HF+BCS method. Extensions in progress involve 
the Hartree-Fock-Bogoliubov (HFB) method and the cranked HFB method. 
No assumption is made on the existence of 
a core and of valence orbitals. The collective basis is 
generated by a constraint on the intrinsic axial quadrupole moment.
This limits the model to even spin and parity collective 
structures. The symmetries broken at the level of the mean
field are restored by means of particle number and angular momentum 
projection. This transfers the physical description from the intrinsic 
to the laboratory frame while yielding moments and transitions to be 
compared directly to experimental data. Finally, the dynamical effects 
associated with the selected constraint are incorporated by means
of a diagonalization of the initial Hamiltonian within the 
collective basis (GCM). Other ways to generate collective bases have
already been explored by our group (see for instance Ref.~\cite{Hee01a}), 
by contraints on
monopole, triaxial quadrupole, and octupole deformations or by the
construction of a two-quasiparticle excited basis. Essentially, these
extensions do not entail a modification of the numerical tools but require 
more computing time and they have only included the restoration of the
particle-number symmetry.

A first test study on the magic nucleus $^{16}$O presented in 
Ref.\ \cite{Ben02a} yielded results demonstrating that the extension of 
a MF method  through restoration of broken symmetries results in a 
significantly improved description of the level structure and other 
spectroscopic properties of this nucleus. The present work
studies four light nuclides -- \Sd, \Ars, \Are, \Caz\ -- to test the 
model on systems illustrative of a variety of properties
encountered in the $s$-$d$ shell. Another reason to select these
nuclei is the recent discovery of superdeformed structures in the 
three heaviest. Such structures are generally interpreted as 
resulting from the interaction between $s$-$d$ and $f$-$p$ shells.

The results presented in this work are encouraging. The general 
trends seen for low-lying collective levels are well described 
both with regard to the relative energy positions within the spectrum 
of a given isotope and to the evolution from nucleus to nucleus.
In particular, as in data, the magic and thus isolated nature of 
the ground state of \Caz\ and \Are\ contrasts with \Ars\ and \Sd, 
where the ground state is strongly coupled to other components 
of the spectrum. Moreover, at small excitation energies, the 
vibrational features of \Sd\ are well reproduced. Within the same 
calculational frame, we also find the recently discovered SD 
structures. We confirm the conclusions drawn from the cranking model
and the shell model 
that the existence of SD bands is connected to a partial occupation 
of deformation-driving orbitals originating from the $f_{7/2}$ shell. 
However, in the specific case of \Caz, a detailed analysis has 
shown that the association of SD bands with 4p-4h configurations must not 
be taken too literally. Quadrupole collectivity generating a coherent 
multi-excitation of p-h configurations of various complexities is 
probably more representative of the nature of the SD bandhead 
wave functions. 
Finally, the $E2$ transition matrix elements of these SD 
bands agree rather well with the data, indicating that our
method correctly reproduces the average quadrupole deformation.
We recall that no effective charge is involved in our calculation.

Our calculations display a systematic deficiency in that they predict
level densities which are too small. This feature manifests itself through 
the high excitation energy of the bandheads and the small values of 
the moments of inertia. At this point, we can only surmise the causes 
of this problem, as it may be traced either to the model, or to 
the Hamiltonian, or to both. However, two results seem to exonerate 
the Hamiltonian, at least as the dominant source of discrepancy. 
First, the results for the magic nucleus $^{16}$O were rather good 
and, just as in this work, angular momentum projection was
shown to lead to large energy gains by amounts of the order of 
5 MeV, reducing by almost an order of magnitude the
MF excitation energies of the deformed configurations
(see Figs.~\ref{F03}, \ref{F07}, \ref{F10} and \ref{F13}).
Second, similar calculations with another well-tested effective 
interaction, i.e.\ the Gogny force, lead to the same phenomenon
\cite{Rod00a}. In this and earlier references from the same
authors, a scaling factor $(\approx0.7)$ is introduced
to compensate for the too low level density. It has been justified on
the basis of a comparison between solutions of a cranked mean-field
method and a projected method on non-cranked states.
In other words, the deficiency does not originate either from 
the Hamiltonian, nor from the model, but from a too strict limitation 
of the MF basis. The best description of the 
$(N_0, Z_0)$ nuclide is obtained by projecting out 
the $(N_0, Z_0)$ component of the MF solution constrained to have the
same mean values of the $\hat{N}$ and $\hat{Z}$ operators. In the same 
way, the optimal description of a state of spin $J$ should 
result from the projection of a cranked MF state constrained to
have \mbox{$\langle {\hat J}_x\rangle = J$}. These points
are also discussed in \cite{RSc80}.
One underlying dynamical explanation has to do with the antipairing 
cranking effect, which increases the moment of inertia as needed to 
improve our results. In the present work, by projecting out a 
\mbox{$J=6$} component out of a non angular momentum MF solution, 
we miss the pairing reduction which affects a 
\mbox{$\langle {\hat J}_x\rangle=6$} cranked MF state.

This discussion also delineates possible lines of future developments.
On the one hand, in selected cases suggested by experimental data, it 
may become necessary to enlarge the collective space by taking into
account other deformations (as suggested, for instance, by Yamagami and
Matsuyanagi~\cite{Yam00a}). Finally, a promising extension of the 
present work would involve the projection of angular momentum cranked 
solutions. Work is presently underway along these three directions.
%
%
\begin{acknowledgments} 
This research was supported in part by the PAI-P5-07 of the Belgian
Office for Scientific Policy. We thank R.~V.~F.\ Janssens and
R.\ Wyss for fruitful and inspiring discussions.  M.~B.\ acknowledges 
support through a European Community Marie Curie Fellowship.
\end{acknowledgments}
%
%
\begin{appendix}
\section*{The quadrupole moment in the lab frame}
In this appendix, we derive the formulas which permits to calculate
quadrupole transition matrix elements between symmetry restored mean-field 
states. Some of the equations given in Ref.~\cite{Val00a} contain misprints 
that are corrected here.

For sake of simple notation, we omit particle-number projection 
throughout this appendix. It does not alter any of the formulas given 
here concerning angular momentum. Different from above, we will 
explicitely include the angular momentum projection $M$ in the notation.
Starting from time-reversal, parity, and axially symmetric (with the
$z$ axis as symmetry axis) mean-field states $| q \rangle$ obtained 
with a constraint on the quadrupole moment labelled as $q$, projected 
states $| J M q \rangle$ with total angular momentum $J$ and angular 
momentum projection $M$
\begin{equation}
| J M q \rangle
= \frac{1}{\mathcal{N}_{Jq}} \hat{P}_{M 0}^{J} | q \rangle
,
\end{equation}
where $\mathcal{N}_{Jq} = \langle q | \hat{P}_{00}^J | q \rangle^{1/2}$
is a normalization factor, are obtained applying the angular-momentum
projector \cite{RSc80}
\begin{equation}
\hat{P}_{M K}^{J}
= \frac{2J+1}{8 \pi^2}
  \int \! d\Omega \; {\mathcal{D}}^{J*}_{MK} (\Omega) \, \hat{R} (\Omega)
,
\end{equation}
where 
$\hat{R} (\Omega) 
= e^{-i \alpha \hat{J}_z} \,
  e^{-i \beta  \hat{J}_y} \,
  e^{-i \gamma \hat{J}_z}
$ 
is the rotation operator and 
$
{\mathcal{D}}^{J}_{MK} (\Omega) 
= e^{-i M \alpha} \, d^J_{MK} (\beta) \, e^{-i K \gamma} 
$
a Wigner function, which depend both on the Euler angles 
$\Omega=(\alpha,\beta,\gamma)$, see \cite{Var88a} for details.

The projected GCM states $| J M k \rangle$ are given by
\begin{equation}
| J M k \rangle
= \int \! dq \; f_{J,k} (q) | J M q \rangle
,
\end{equation}
where the non-orthonormal set of weight functions $f_{J,k} (q)$ is 
determined variationally by solving the Hill-Wheeler equation. The 
collective wave functions $g_{J,k} (q)$ given in figures \ref{F05}, 
\ref{F09}, \ref{F12} and \ref{F15} above are obtained from the $f_{J,k} (q)$ 
by a transformation involving the hermitean square root of the norm 
operator, see \cite{BHR03} and references therein for more details.

The spectroscopic quadrupole moment of the GCM state $|J M k \rangle$ 
is defined as
\begin{eqnarray}
\label{eq:Qs}
Q_c (J,k)
& = & \sqrt{\frac{16\pi}{5}} \,
      \langle J, M=J, k | \hat{Q}_{2 0} | J, M=J, k \rangle
      \nn \\
& = & \sqrt{\frac{16\pi}{5}} \,
      \int \! dq \int \! dq' \; f_{J,k}^{*} (q ) \, 
                                f_{J,k}     (q') 
      \nn \\
&   & \quad \times
      \langle J, M=J, q | \hat{Q}_{2 0} | J, M=J, q' \rangle
.
\end{eqnarray}
For $J$ and $J'$ integer and even -- as assumed here -- the matrix 
element of the projected mean-field states entering Eq.\ (\ref{eq:Qs}) 
can be expressed in terms of the reduced matrix element 
$\langle J q || \hat{Q}_{2 0} || J q' \rangle$ \cite{Var88a}
\begin{eqnarray}
\lefteqn{
\langle J, M=J, q | \hat{Q}_{2 0} | J, M=J, q' \rangle
}
      \nn \\
& \quad & = \frac{\langle J J 2 0 | J J \rangle}{\sqrt{2J+1}}
            \langle J q || \hat{Q}_{2 0} || J q' \rangle
.
\end{eqnarray}
The reduced matrix element involving different values of $J$ appears in 
the expression of the reduced transition probability between the GCM 
states $k'$ and $k$ with angular momentum $J'$ and $J$, respectively
\begin{widetext}
\begin{eqnarray}
B (E2, J_{k'}' \to J_k)
& = & \frac{e^2}{2J'+1} 
      \sum_{M =-J }^{+J }
      \sum_{M'=-J'}^{+J'}
      \sum_{\mu=-2}^{+2}
      | \langle J M k | \hat{Q}_{2 \mu} | J' M' k' \rangle |^2
      \nn \\
& = & \frac{e^2}{2J'+1} 
      \left| \int \! dq \int \! dq' \; f^{*}_{J,k} (q) \,
      f_{J',k'} (q') \,
      \langle J q || \hat{Q}_{2} || J' q' \rangle
      \right|^2
.
\end{eqnarray}
\end{widetext}
The reduced matrix element is evaluated by the commutation of
one of the two projection operators in the expression
$
\hat{P}_{KM}^J \hat{Q}_{2 \mu} \hat{P}_{M'K'}^{J'}
$
with the quadrupole operator. After a lengthy, but straightforward 
application of angular momentum algebra and taking advantage of the 
symmetries of the mean-field states, one obtains
\begin{widetext}
\begin{eqnarray}
\langle J q || \hat{Q}_{2} || J' q' \rangle
& = & \frac{\sqrt{2J+1} \, (2J'+1)}
      {\mathcal{N}_{Jq} \, \mathcal{N}_{J'q'}}
      \sum_{\mu=-2}^{+2} 
      \frac{1+(-)^{J}}{2}
      \langle J' 0 2 \mu | J \mu' \rangle 
      \int \limits_{0}^{\pi/2} \! d\beta \; \sin (\beta) \, 
      d^{J}_{0 \mu} (\beta) \, 
      \langle q | e^{-i \beta \hat{J}_y} \, \hat{Q}_{2 \mu} | q' \rangle
\end{eqnarray}
\end{widetext}
where only the ``left'' states have to be rotated. A similar expression, 
where only the right states have to be rotated, is given in \cite{Rod02a}.
\end{appendix}
%
%

\end{document}